\documentclass[aps,pra,superscriptaddress,longbibliography,showpacs,twocolumn,10pt]{revtex4-1}

\usepackage{graphicx}
\usepackage{epstopdf}
\usepackage{amsmath}
\usepackage{amsfonts}
\usepackage{amssymb}
\usepackage{hyperref}

\begin{document}


\title{Conductance oscillations in quantum point contacts of InAs/GaSb heterostructures}
\author{Micha{\l} Papaj}
\altaffiliation{present address: Department of Physics,
Massachusetts Institute of Technology, Cambridge, MA 02139, USA \\ e-mail: mpapaj@mit.edu}
\affiliation{Institute of Theoretical Physics, Faculty of Physics, University of Warsaw,
ulica Pasteura 5, PL-02-093 Warszawa, Poland}

\author{{\L}ukasz Cywi\'nski}
\email[]{lcyw@ifpan.edu.pl}
\affiliation{Institute of Physics, Polish Academy of Sciences,
aleja Lotnik\'{o}w 32/46, PL-02-668 Warszawa, Poland}

\author{Jerzy Wr\'obel}
\affiliation{Institute of Physics, Polish Academy of Sciences,
aleja Lotnik\'{o}w 32/46, PL-02-668 Warszawa, Poland}
\affiliation{Faculty of Mathematics and Natural Sciences, Rzesz\'ow University,
aleja Rejtana 16A, PL-35-959 Rzesz\'ow, Poland}

\author{Tomasz Dietl}
\affiliation{Institute of Theoretical Physics, Faculty of Physics, University of Warsaw,
ulica Pasteura 5, PL-02-093 Warszawa, Poland}
\affiliation{Institute of Physics, Polish Academy of Sciences,
aleja Lotnik\'{o}w 32/46, PL-02-668 Warszawa, Poland}
\affiliation{WPI-Advanced Institute for Materials Research (WPI-AIMR),
Tohoku University, 2-1-1 Katahira, Aoba-ku, Sendai 980-8577, Japan}




\date{\today}

\begin{abstract}
We study quantum point contacts in two-dimensional topological insulators by means of quantum transport simulations for InAs/GaSb heterostructures and HgTe/(Hg,Cd)Te quantum wells. In InAs/GaSb, the density of edge states shows an oscillatory decay as a function of the distance to the edge. This is in contrast to the behavior of the edge states in HgTe quantum wells, which decay into the bulk in a simple exponential manner. The difference between the two materials is brought about by spatial separation of electrons and holes in InAs/GaSb, which affects the magnitudes of the parameters describing the particle-hole asymmetry and the strength of intersubband coupling within the Bernevig-Hughes-Zhang model. We show that the character of the wave function decay impacts directly the dependence of the point contact conductance on the constriction width and the Fermi energy, which can be verified experimentally and serve to determine accurately the values of relevant parameters. In the case of InAs/GaSb heterostructures the conductance magnitude oscillates as a function of the constriction width following the oscillations of the edge state penetration, whereas in HgTe/(Hg,Cd)Te quantum wells a single switching from transmitting to reflecting contact is predicted.
\end{abstract}

\pacs{}

\maketitle

\section{Introduction}
The quantum spin Hall effect (QSHE) is characterized by a coexistence of insulating bulk and conducting helical edge modes that are topologically protected from backscattering \cite{hasan_colloquium:_2010, qi_topological_2011, dolcetto_edge_2016}. So far, QSHE has been predicted and observed for HgTe/(Hg,Cd)Te quantum wells (QWs) \cite{bernevig_quantum_2006, konig_quantum_2007, roth_nonlocal_2009} and InAs/GaSb heterostructures \cite{liu_quantum_2008, knez_evidence_2011, du_robust_2015, li_observation_2015}. However, the accuracy of the conductance quantization in these systems is substantially inferior compared to the case of the quantum Hall effect \cite{schopfer_quantum_2013} and the quantum anomalous Hall effect \cite{chang_experimental_2013}. For instance, in the case of HgTe/(Hg,Cd)Te QWs, while non-local transport studies clearly indicate the existence of the edge conducting channels \cite{gusev_nonlocal_2012, grabecki_nonlocal_2013}, the magnitude of conductance approaches the theoretically expected quantized values only in micrometer-sized structures, i.e., when the distance between the probes is of the order of the mean free path \cite{konig_quantum_2007, roth_nonlocal_2009,olshanetsky_persistence_2015}. Furthermore, surprising aperiodic and reproducible conductance fluctuations are reducing the accuracy of quantization even further \cite{konig_quantum_2007, roth_nonlocal_2009, gusev_nonlocal_2012, grabecki_nonlocal_2013, olshanetsky_persistence_2015}. 
Finally, recently it has been suggested that topologically trivial edge states coexist with the helical ones both in HgTe/(Hg,Cd)Te QWs \cite{Ma_NC15} and in InAs/GaSb heterostructures \cite{Nichele_arXiv15}.
This situation calls for identification of experimental scenarios in which the helical nature of edge states leads to appearance of characteristic phenomena other than the approximate level of conductance agreeing with the simplest theoretical predictions.

Understanding these discrepancies between theory and experiment requires further studies on the nature of the edge states in 2D TIs and their interaction with the environment. There are theoretical proposals of both intrinsic \cite{kane_$z_2$_2005, xu_stability_2006, wu_helical_2006, schmidt_inelastic_2012, budich_phonon-induced_2012, maciejko_kondo_2009, tanaka_conductance_2011, lunde_helical_2012} and extrinsic \cite{girschik_topological_2013, vayrynen_helical_2013, vayrynen_resistance_2014, essert_magnetotransport_2015, essert_two-dimensional_2015, grabecki_nonlocal_2013} mechanisms responsible for the imperfect conductance quantization. One of the important areas of research is concerned with the coupling of the states localized on the opposite edges, which can lead to backscattering. Experimentally, controllable interactions of the opposite edges can be realized by using quantum point contacts (QPC). QPCs are nanoconstrictions introduced into the samples via nanofabrication with side gates or back gates employed for controlling the width  \cite{van_wees_quantized_1988, van_wees_quantum_1991} or diameter \cite{Grabecki:2002_PE} of the point contact. So far, in the context of 2D TIs quantum point contacts have been theoretically studied either for HgTe QWs \cite{krueckl_switching_2011, zhang_electrical_2011, fu_perfect_2014} or for general models of helical edge states \cite{dolcini_full_2011, romeo_electrical_2012, orth_point_2013, ferraro_electronic_2014}. Apart from their usefulness in studying the size effects and interedge scattering, various possible applications of QPCs prepared from QSHE materials include: (i) usage of interferometry to control spin and charge conductances \cite{dolcini_full_2011, romeo_electrical_2012}, (ii) studying localization effects of helical edge states \cite{orth_point_2013}, and (iii) performing Hong-Ou-Mandel-type experiments \cite{ferraro_electronic_2014}. Hence, QPC structures constitute the basic building blocks of electron optics. At the same time, QPC can be employed to control electrical current in the devices of 2D topological insulators by closing and opening the constriction by side gates \cite{krueckl_switching_2011, zhang_electrical_2011, fu_perfect_2014}.

The possibilities outlined above have motivated our conductance study of point contacts prepared from both classes of materials that exhibit QSHE. In particular, we report on quantum transport simulations for QPCs patterned from either InAs/GaSb heterostructures or HgTe/(Hg,Cd)Te QW. Surprisingly, our results point to quite a different behavior of the edge state decay as a function of the distance to the edge in these two materials - a standard exponential decay in the case of HgTe but an oscillating behavior in InAs/GaSb. We have checked that this striking difference is beyond uncertainties in parameters values as well is immune to disorder and to supplementing the Hamiltonian by Dresselhaus and Rashba terms. Our finding impacts directly the conductance of the QPC as a function of the channel width or side gate voltage. A single cross-over from conducting to insulating state is predicted for HgTe QW, whereas multiple switching between the two states is expected for nanoconstriction of InAs/GaSb heterostructures. These features can be verified experimentally, and would provide direct information on the character of decay of the helical edge states towards QW center. 
Furthermore, the extent of the edge wave function determines directly the strength of carrier-mediated ferromagnetic coupling between localized spins \cite{Dietl:2014_RMP}, a key characteristics of (Hg,Mn)Te/(Hg,Cd)Te QWs and related systems \cite{Liu:2009_PRL}.

\section{Methods}
Results presented in this work have been obtained by using the Bernevig-Hughes-Zhang Hamiltonian \cite{bernevig_quantum_2006}. It acts on wavefunctions of the form $\psi = (\psi_{s\uparrow}, \psi_{p\uparrow}, \psi_{s\downarrow}, \psi_{p\downarrow})$ ($s$ and $p$ denote different subbands, while $\uparrow$ and $\downarrow$ account for spin projection) and has the form,

\begin{equation}
\label{eq:BHZ_hamiltonian}
\mathcal{H} (\mathbf{k}) = \begin{pmatrix} H_0(\mathbf{k}) & H_1(\mathbf{k}) \\ -H_1^*(-\mathbf{k}) & H_0^*(-\mathbf{k})\end{pmatrix}
\end{equation}

\begin{equation}
H_0(\mathbf{k}) = \begin{pmatrix}M-(B+D)k^2 & A k_+ \\ Ak_- & -M+(B-D)k^2\end{pmatrix}
\end{equation}

\begin{equation}
H_1(\mathbf{k}) = \begin{pmatrix}\Delta_e k_+ + i\xi_e k_- & - \Delta_z \\ \Delta_z & \Delta_h k_-\end{pmatrix}
\end{equation}
where $\mathbf{k} = (k_x, k_y)$, $k^2 = k_x^2 + k_y^2$, $k_\pm = k_x \pm i k_y$. $H_0 (\mathbf{k})$ blocks of the matrix are responsible for the intersubband coupling, while $H_1 (\mathbf{k})$ consists of terms responsible for Dresselhaus ($\Delta_e k_+$, $\Delta_h k_-$ and $\Delta_z$) \cite{liu_quantum_2008} and Rashba ($i \xi_e k_-$) \cite{rothe_fingerprint_2010} spin-orbit coupling. The values of parameters used in the simulations are collected in Table \ref{tab:parameters} \cite{franz_topological_2013}. As will be shown below, the main differences between both materials are in the particle-hole symmetry term $D$ and the intersubband coupling $A$.

\begin{table}
\caption{\label{tab:parameters} Parameters of the tight-binding Hamiltonian \ref{eq:BHZ_hamiltonian} for 7\,nm thick HgTe/(Hg,Cd)Te quantum well and 10\,nm/10\,nm thick InAs/GaSb heterostructure from Ref.~\onlinecite{franz_topological_2013}.}
\begin{ruledtabular}
\begin{tabular}{lcc}
Parameter & HgTe QW & InAs/GaSb heterostructure \\
\hline \hline
A [eV\,{\AA}] & 3.65 & 0.37 \\
B [ eV\,{\AA}$^2$] & -68.6 & -66.0 \\
D [ eV\,{\AA}$^2$] & -51.1 & -5.8 \\
M [ eV] & -0.01 & -0.0078 \\
$\Delta_z$ [ eV] & 0.0016 & 0.0002 \\
$\Delta_e$ [ eV\,{\AA}] & -0.128 & 0.00066 \\
$\Delta_h$ [ eV\,{\AA}] & 0.211 & 0.0006 \\
$\xi_e$ [ eV\,{\AA}] & 0.0 & -0.07 \\
\end{tabular}
\end{ruledtabular}
\end{table}

\begin{figure}
\includegraphics[width=0.49\textwidth]{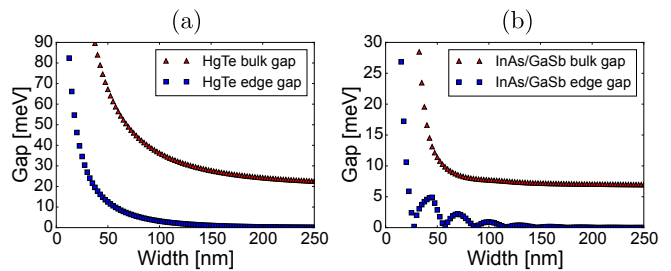}
\caption{\label{fig:band_gap} (Color online) Dependence of the bulk and edge state band gaps on the bar width for (a) HgTe QW (b) InAs/GaSb heterostructure.}
\end{figure}

\begin{figure}
\includegraphics[width=0.48\textwidth]{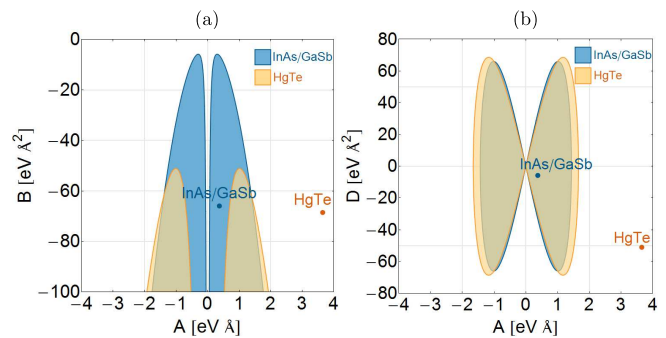}
\caption{\label{fig:inequality} (Color online) Regions in the parameter space of the BHZ model for which edge state oscillations occur in the case of InAs/GaSb (blue) and HgTe (orange) for $E$=0 and varying (a) A and B parameters, (b) A and D parameters. The points show the parameters used in the simulations.}
\end{figure}

\begin{figure*}
\includegraphics[width=\textwidth]{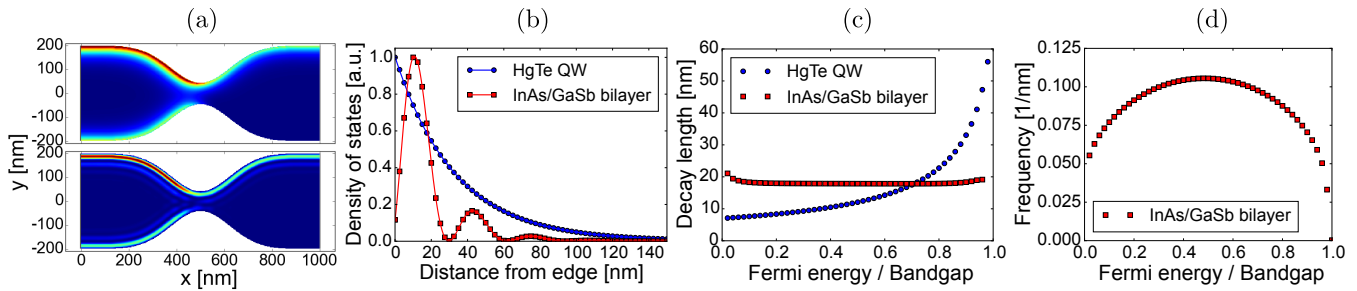}
\caption{\label{fig:edge_states} (Color online) (a) Wavefunction of spin up electrons entering the sample from the left lead (upper: HgTe QW, lower: InAs/GaSb heterostructure). (b) Local density of states near the edge of HgTe QW and InAs/GaSb heterostructure bars with corresponding fitted curves. HgTe density of states follows simple exponential decay, while InAs/GaSb heterostructures show oscillatory decay. (c) Decay length of the edge states across the bulk band gap. Values are obtained through fitting to the density of states and are constant for InAs/GaSb and monotonically growing for HgTe. (d) Frequency of the density of states oscillations across the band gap.}
\end{figure*}

The Hamiltonian \ref{eq:BHZ_hamiltonian} has been discretized on a square lattice with lattice constant $a=2.5$\,nm. The area under study has width $W = 400$\,nm and length $L = 1000$\,nm. To this rectangular area we attach semi-infinite ballistic leads made of the same material. For simulations of a QPC in the middle of this area we place a constriction that has a shape of two identical Gaussian functions $f(x) = F \exp(-(x-L/2)^2/2\sigma^2)$ placed on the opposite edges of the sample [see the wavefunction map in Fig.~\ref{fig:edge_states}(a)]. In all further considerations we define the width of the constriction as the distance between the maxima of the Gaussian functions, that is $W-2F$ and the length of the constriction is defined as $\sigma$. All the numerical calculations have been performed using the Kwant simulation toolbox \cite{groth_kwant:_2014}. In all conductance calculations we determine the scattering matrix (and subsequently the transmission matrix $t$) and then employ Landauer-B\"uttiker formula, $G = e^2/h \,\mathrm{Tr}\,t^\dagger t$. In some of the calculations, spin-independent disorder is added in a form of a potential $U(\mathbf{r})$ that enters the Hamiltonian in the same way as $M$ parameter, modifying the onsite energies. The disorder energy values are uncorrelated between the sites and taken from a random distribution in the range $[-V/2, V/2]$ with $V$ being the disorder strength. The simulations with disorder are averaged over a number of independent disorder realizations, typically of the order of 50.

\section{Results}
For the parameters from Table \ref{tab:parameters}, we calculate the band structures of 1D ribbons with various widths. Due to a finite width of the constriction, in addition to the bulk band gap there is a gap for the edge states, which arises due to coupling of states from the opposite sides of the sample \cite{zhou_finite_2008}. In both studied materials the bulk band gap monotonically decreases (Fig.~\ref{fig:band_gap}) and for $W = 400$\,nm assumes the value of 20.8\,meV and 6.8\,meV for HgTe QW and InAs/GaSb heterostructures, respectively. However, the behavior of the gap for edge states is different in the two materials. The edge gap in HgTe QW decays approximately exponentially to zero, while in InAs/GaSb heterostructures the magnitude of the gap oscillates, for some widths dropping to zero and increasing again back to the value of several meV. This difference in the width dependence of the gap value coincides with a disparate decay character of the edge states in both materials. For energies inside the bulk band gap the local density of states (LDOS) near the edge is shown in Fig.~\ref{fig:edge_states}(b). While the edge states of HgTe quantum wells follow a simple exponential decay, the decay of LDOS in InAs/GaSb heterostructures shows an oscillatory pattern. This can be understood by considering the analytic solution for the edge states in the BHZ model without additional Dresselhaus and Rashba terms \cite{zhou_finite_2008}. The decay constant for the simplified model becomes complex for parameters that satisfy the inequality,
\begin{equation}
\label{eq:inequality}
\frac{A^4-4 A^2 (D E+B M)+4 (B E+D M)^2}{4 \left(B^2-D^2\right)^2}<0
\end{equation}
where $E$ is the energy eigenvalue for which the Hamiltonian is solved. Examples of regions in parameter space for which the oscillations occur are presented in Fig.~\ref{fig:inequality}. The colored areas are obtained by varying $A$ and $B$ [Fig.~\ref{fig:inequality}(a)], and $A$ and $D$ [Fig.~\ref{fig:inequality}(b)], and holding the remaining parameters constant at appropriate values for both materials. The points inserted in the figures show the parameter values used for the simulations. The set of parameters corresponding to HgTe QWs lies well outside the region of oscillations for HgTe, while the point corresponding to InAs/GaSb resides in the middle of the blue region of oscillations for this material.

When the Dresslehaus and Rashba terms are introduced into the Hamiltonian, the oscillations can be obtained in the case of HgTe QW, as noted previously \cite{takagaki_backscattering_2012}. However, this effect is of a negligible magnitude and requires a substantial increase in the strength of the Dresselhaus term to be observable for sample widths smaller than 200\,nm and to have an amplitude above 1\,meV \cite{zhi_finite_2014, takagaki_cancelation_2014}. On the other hand, in the case of InAs/GaSb even without the spin-orbit terms in the Hamiltonian the oscillation is present and comparable to the result presented in Fig.~\ref{fig:band_gap}(b). There is no parameter fine-tuning required as the region determined from inequality \ref{eq:inequality} is sufficiently broad. For example, the effect is also present when the calculations are performed using the values for 10\,nm/9\,nm thick InAs/GaSb heterostructure \cite{franz_topological_2013}. Furthermore, the magnitude of the effect is sufficiently large to have a direct impact on the transport properties of the quantum point contacts, which is a feature that has not been considered previously.

As the HgTe QW has structural inversion symmetry, the $\xi_e$ term vanishes in the absence of an external electric field. Even though such a field is present when external gate is used to shift the Fermi level, because the influence of the Rashba term is even smaller than that of the Dresselhaus contribution \cite{zhi_finite_2014}, the behavior of the edge states does not change drastically for $e\mathcal{E}_z$ up to $10^7$\,eV/m. Therefore in subsequent calculations the impact of perpendicular electric field on spin-orbit coupling terms is neglected.

To determine the decay parameters in the case of the model with Rashba and Dresselhaus spin-orbit coupling, we fit curves to LDOS calculated for various energies inside the band gap. In the case of HgTe the fitting function is $P \exp(-x/l_0)$ and for InAs/GaSb it has a form $P \exp(-x/l_0) \sin^2 (fx+c)$, where $l_0$ is the decay length of the edge states and $f$ is the frequency of LDOS oscillations. The obtained decay lengths for all energies inside the band gaps are shown in Fig.~\ref{fig:edge_states}(c). The decay length of the edge states in HgTe QW grows monotonically across the band gap and increases sixfold from the top of the bulk valence band to the bottom of the bulk conduction band.  On the other hand, the decay length of edge states in InAs/GaSb heterostructures remains approximately constant across the band gap. This difference is due to large particle-hole asymmetry term used to parametrize the HgTe quantum well - for a reduced value of the parameter $D$ the decay length changes only by a few nanometers in HgTe QW, too. The difference between the decay character of the edge states is important for the properties of conductance in quantum point contacts and again it is a consequence of coupling of electron and hole bands. Another value retrieved from the fitting procedure in the case of heterostructures is the frequency of LDOS oscillations, whose dependence on the position of the Fermi level inside the band gap is shown in Fig.~\ref{fig:edge_states}(d). This frequency is close to zero in the vicinity of the conduction band edge, and in the middle of the gap it reaches its maximum value, which is twice as high as the value attained close to the valence band.

\begin{figure*}
\includegraphics[width=0.85\textwidth]{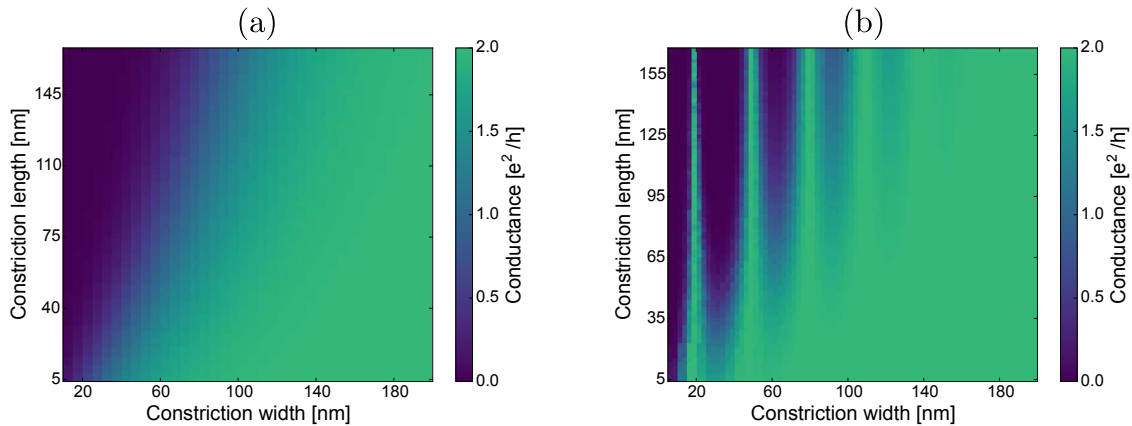}
\caption{\label{fig:length_width} (Color online) Conductance of the Gaussian-shaped quantum point contact in (a) HgTe quantum well, (b) InAs/GaSb heterostructure for varying dimensions of the constriction and material parameters from Table \ref{tab:parameters}. The Fermi levels in both materials are placed close to the Dirac point at 7\,meV and 0.6\,meV, respectively.}
\end{figure*}

In Fig.~\ref{fig:length_width} we present the results of quantum transport simulations for quantum point contacts of variable length and width for both HgTe QW and InAs/GaSb heterostructure. For both materials the Fermi level is placed close to the Dirac point at 7\,meV and 0.6\,meV, respectively. For HgTe QW (Fig.~\ref{fig:length_width}a) we see a smooth transition between transmitting and reflecting quantum point contact. The constriction width, for which the QPC closes, increases with the constriction length. As we study coherent transport, closing of the QPC is only due to mixing of the edge states from the opposite edges of the sample. Those new hybridized states are gapped and non-helical and in their case backscattering is allowed. As the edge states decay exponentially in HgTe QW, the closer the opposite edges are, the greater the probability of edge state mixing and thus interedge scattering. This probability increases also with the length of the constriction, because the region in which both sides of the sample are close to each other is larger. On the other hand, in the InAs/GaSb heterostructures the edge states decay with an oscillatory pattern and so for some edge separations the overlap of wavefunctions is minimized. This is illustrated in Fig.~\ref{fig:length_width}(b), where conductance decreases and increases in the oscillatory pattern that has the same frequency as the oscillations of the density of states.

\begin{figure*}
\includegraphics[width=0.85\textwidth]{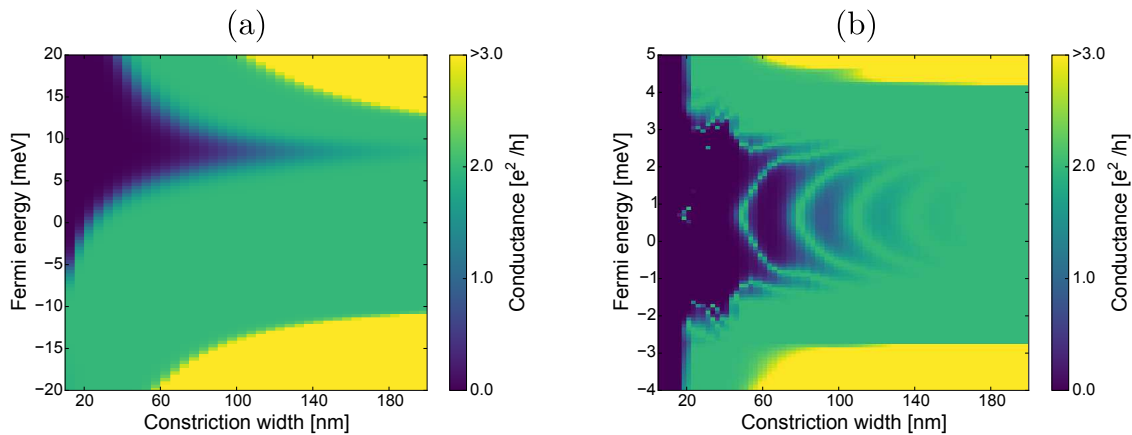}
\caption{\label{fig:fermi_width} (Color online) Conductance of the Gaussian-shaped quantum point contact in (a) HgTe quantum well, (b) InAs/GaSb heterostructure for varying Fermi energies and constriction widths and material parameters from Table \ref{tab:parameters}. In both contacts the length of the constriction is set to 168\,nm.}
\end{figure*}

In Fig.~\ref{fig:fermi_width} the conductance maps for varying Fermi energies and constrictions widths are shown. For HgTe QW [Fig.~\ref{fig:fermi_width}(a)] we observe two regions inside the band gap: the one closer to the valence band, in which QPC perfectly transmits electrons for large range of constrictions widths and the region closer to the conduction band, where backscattering and thus reflection of electrons occurs even for large edge separations. This is consistent with the result presented in Fig.~\ref{fig:edge_states}(c): for Fermi energies closer to the conduction band the edge states penetrate deeper into the sample and so have a higher chance to scatter to the states at the opposite edge for the same constriction width. In the case of InAs/GaSb heterostructure [Fig.~\ref{fig:fermi_width}(b)] again the oscillatory decay of the edge states impacts the conductance. There are rings of perfect conductance visible in the map and their separation coincides with the separation of nodes of the density of states. As the frequency of LDOS oscillations changes throughout the band gap [Fig.~\ref{fig:edge_states}(d)], the separation of the rings changes, too. This also confirms the connection between the decay character and QPC conductance.

One should note that the simulation results presented above do not include effects of electron-electron interactions. In quantum point contacts the increased overlap of edge channels can lead not only to the enhanced backscattering but also to the Coulomb blockade \cite{romeo_interaction_2014}. This can obscure the features observed in our simulations by significantly reducing the conductance of the QPC. However, for the InAs/GaSb devices the overlap in the constriction oscillates as a function of the width and so the strength of the Coulomb blockade effect will be dependent on the edge separation. This in turn means that the reduction of conductance will be greater for the conductance minima than for the maxima and, thus, the relative amplitude of the oscillations will be larger. Also, the calculations \cite{romeo_interaction_2014} show that the impact of the Coulomb blockade can be minimized by an appropriate interplay of bias and side gate voltages.

\begin{figure*}
\includegraphics[width=0.7\textwidth]{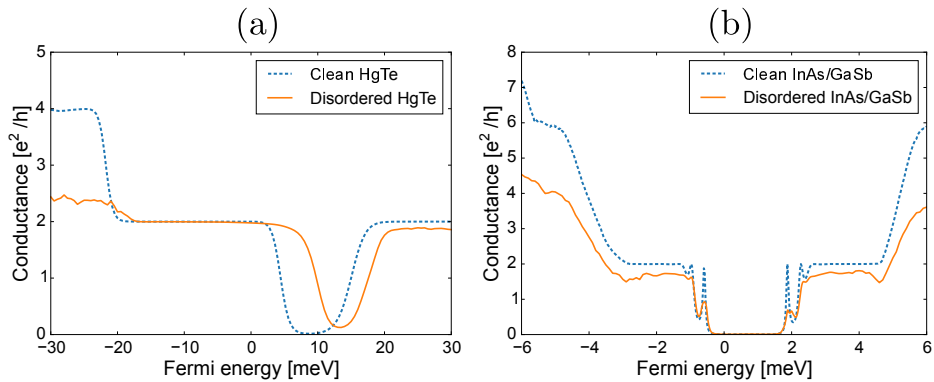} \quad
\caption{\label{fig:switching} (Color online) Conductance of the quantum point contact for: (a) HgTe QW (constriction width 55\,nm and length 168\,nm), (b) InAs/GaSb heterostructure (constriction width 65\,nm and length 212\,nm) for clean and disordered systems (averaged over 50 disorder realizations).}
\end{figure*}

Experimentally, the Fermi energy can be shifted using a top gate, and  this feature can be employed in devices as an edge state current switch \cite{krueckl_switching_2011}. In Fig.~\ref{fig:switching} we show dependence of the QPC conductance on the position of the Fermi level for HgTe QW [\ref{fig:switching}(a), constriction width 55\,nm and length 168\,nm] and for InAs/GaSb heterostructure [\ref{fig:switching}(b), constriction width 65\,nm and length 212\,nm], both for a clean and disordered systems. While for the clean systems it is possible to obtain less than 1\% of perfect $2e^2/h$ conductance in the "off" state of the quantum point contact and $2e^2/h$ with accuracy of 10$^{-5}$ in the "on" state in both cases, we observe that both systems display different behavior in the presence of disorder (strengths $V = 125$\,meV for HgTe and $V = 30$\,meV for InAs/GaSb). In HgTe QWs, the conductance of the edge states retains a value close to the perfect one for disorder strength that breaks down the conduction of the bulk modes in the valence band, and the QPC can still function as a current switch. However, the edge current in the InAs/GaSb case is much less protected, and even for smaller disorder strength the conductance decreases significantly below $2e^2/h$. However, the oscillations in conductance are still visible and therefore they can still be experimentally detected.

\section{Conclusion}
In summary, we have shown how the strength of intersubband coupling in 2D topological insulators impacts the behavior of the edge states in samples of finite width, and in consequence the conductance in such systems. In HgTe/(Hg,Cd)Te quantum wells, a system with strong electron-hole coupling, the edge states follow a simple exponential decay, while in InAs/GaSb heterostructures, where the coupling is much weaker, the edge states develop an oscillatory decay pattern. This is closely followed by the dependence on the constriction width and the Fermi energy of conductance in quantum point contacts, where the mixing of the edge states from the opposite edges of the sample is emphasized. QPCs in HgTe/(Hg,Cd)Te quantum wells display a single transition between the transmitting and reflecting state, whereas in InAs/GaSb the state of the contact switches periodically. We have also shown that this difference between the two classes of 2D topological insulators may be observed in the presence of disorder.
The oscillatory behavior of conductance of quantum point contact made out of the InAs/GaSb heterostructure tuned into a topologically nontrivial regime could be thus considered a fingerprint of presence of helical edge states.

\begin{acknowledgments}
This work was supported by the Polish Ministry of Science and Higher Education "Diamentowy Grant" Project No. DI2013 016243 and in part by the National
Center of Science in Poland (Decision No.\,2011/02/A/ST3/00125).
\end{acknowledgments}

\bibliography{ti_qpc}

\end{document}